\documentclass[aps,11pt,nofootinbib,preprintnumbers,showpacs]{revtex4}

\usepackage{amssymb, amsfonts, amsmath, bm}
\usepackage{amsfonts}
\usepackage[dvipdfmx]{graphicx,color}

\begin{document}




\title{
 Charged black lens in de Sitter space 
}
\vspace{2cm}

\author{
Shinya Tomizawa\footnote{tomizawasny`at'stf.teu.ac.jp} 
} 
\vspace{2cm}
\affiliation{
Department of Liberal Arts, Tokyo University of Technology, 5-23-22, Nishikamata, Otaku, Tokyo, 144-8535, Japan
}




\begin{abstract} 
We obtain a charged black lens solution in the five-dimensional Einstein-Maxwell-Chern-Simons theory with a positive cosmological constant. 
It is shown that the solution obtained here describes the formation of a black hole with the spatial cross section of a sphere from that of the lens space of $L(n,1)$ in five-dimensional de Sitter space.
\end{abstract}

\pacs{04.50.+h  04.70.Bw}
\date{\today}
\maketitle




\section{Introduction}\label{sec:1}
Black holes in more than four spacetime dimensions have been one of the most interesting subjects of general relativity in the last two decades. For example, the statistical counting of black-hole entropy was performed in string theory~\cite{Strominger:1996sh}, which needs higher-dimensional gravity.  
Also,  the AdS/CFT correspondence~\cite{Aharony:1999ti} relates the dynamics of higher-dimensional black holes to those of a quantum field theory in one lower dimension. 
Moreover, the higher-dimensional black-hole production at an accelerator was predicted in the scenario of large extra dimensions~\cite{Argyres:1998qn}. 
In spite of a lot of efforts by many physicists, our understanding of higher-dimensional black holes cannot be said to be enough, even in five-dimensional pure gravity.

\medskip

The topology theorems for stationary black holes in five dimensions~\cite{Galloway:2005mf,Cai:2001su,Hollands:2007aj,Hollands:2010qy} state that the allowed topology of the spatial cross section of the event horizon is either a sphere $S^3$, a ring $S^1\times S^2$ or lens spaces $L(p,q)$, if the spacetime is asymptotically flat and admits biaxisymmetry.  
For both the sphere and the ring, the corresponding exact solutions have been found as stationary solutions to the five-dimensional vacuum Einstein  equations~\cite{Tangherlini:1963bw,Myers:1986un,Emparan:2001wn,Pomeransky:2006bd}, whereas for the lens spaces, they have not yet been found, although some authors have tried to find them as regular vacuum solutions~\cite{Evslin:2008gx,Chen:2008fa}. 
Recently, however, as for the class of supersymmetric solutions in the bosonic sector of five-dimensional minimal ungauged supergravity,  the first black lens solution whose horizon topology is $L(2,1)=S^3/{\mathbb Z}_2$ has been constructed  by Kunduri and Lucietti~\cite{Kunduri:2014kja} with the help of the well-known construction developed by Gauntlett {\it et al.}~\cite{Gauntlett:2002nw}.  Moreover, this solution has been extended to those that admit the horizon with the more general lens space topologies $L(n,1)=S^3/{\mathbb Z}_n$~\cite{Tomizawa:2016kjh} and multiple horizons~\cite{Tomizawa:2017suc}.

\medskip
The next step that one should consider may be whether a black lens exists in a de Sitter space and an anti-de Sitter space.  
To find such a solution may seem to be a considerably difficult problem since not only do we not have the solution-generation-methods for the Einstein equations with a cosmological constant, but also we do not know even a (regular) vacuum solution of the black lens.
However, if we start with an extremal black hole solution to the Einstein-Maxwell-Chern-Simons (EMCS) equations, it is not too difficult to find such an exact solution.  
For instance, the Kastor-Traschen (KT) solution, which describes colliding charged black holes for a contracting phase, was found by adding a positive cosmological constant to the Majumdar-Papapetrou (MP) multi-black-hole solutions. 
Applying the same method to higher dimensions, London  has also found the KT solutions in arbitrary higher dimensions that present the coalescence of an arbitrary number of spherical black holes into a single spherical black hole from the MP solutions in arbitrary higher dimensions. 
Moreover, using the method for the rotating case and deforming the well-known supersymmetric Breckenridge-Myers-Peet-Vafa (BMPV) solution~\cite{Breckenridge:1996is} so that it has a cosmological constant, Klemm and Sabra have obtained the five-dimensional rotating KT solution that  presents the coalescence of any number of rotating black holes.

\medskip
The regular metric on the lens space $L(n,1)=S^3/{\mathbb Z}_n$ with unit radius can be written as 
\begin{eqnarray}
ds^2=\frac 14 \left[\left(\frac{d\psi}{n}+\cos\theta d\phi\right)^2+d\theta^2+\sin^2\theta d\phi^2\right] ,
\label{eq:lens}
\end{eqnarray}
where $0\le \psi<4\pi$, $0\le \phi<2\pi$ and $0\le \theta\le\pi$, and $n$ is a positive integer parametrizing the Chern class of the principal bundle over $S^2$. 
In particular, for $n=1$,  this coincides with the metric on a three-dimensional sphere written in terms of the Euler angle coordinates. 
The purpose of this paper is to seek an exact solution of a black lens solution such that the spatial cross section of the horizon can be written in the above form in the five-dimensional EMCS theory with a positive cosmological constant  by adding the found supersymmetric black lens solution to a positive cosmological constant. 
 The method is essentially based on some previous works in Refs.~\cite{Kunduri:2014kja,Tomizawa:2016kjh} by Klemm and Sabra~\cite{Klemm:2000gh}. 
 It is worthy of special mention that while the cosmological black hole with the horizon of $S^3$ in Ref.~\cite{Klemm:2000gh} is stationary, the cosmological black lens solution is dynamical in the sense that the horizon topology changes from the lens space of $L(n,1)=S^3/{\mathbb Z}_n$ into a sphere $S^3$. In fact, using the cosmological chart, we can analytically show that at early time the horizon is isometric to the lens space $L(n,1)=S^3/{\mathbb Z}_n$, whereas at late time it is isometric to a sphere $S^3$. 
For the stationary supersymmetric  black lens solution,  a black lens with zero angular momenta cannot be realized~\cite{Kunduri:2014kja,Tomizawa:2016kjh}, while for the cosmological black lens solution (at least, at early time), it can be realized due to the existence of a cosmological constant.

\medskip

This paper is organized as follows.   
In Sec.~\ref{sec:Review}, we first review the supersymmetric black lens solution~\cite{Tomizawa:2016kjh} in the bosonic sector of five-dimensional minimal supergravity.
Next, we review the Klemm-Sabra (KS) solution of Ref.~\cite{Klemm:2000gh}, which can be regarded as a cosmological BMPV solution.  
In Sec.~\ref{sec:solution}, we explicitly present the cosmological black lens solution in the five-dimensional EMCS theory with a positive cosmological constant.
In Sec.\ref{sec:analysis}, it is shown that---unlike the Klemm-Sabra solution---this solution obtained here describes a dynamical spacetime, which can be seen by showing  that
the horizon topology changes from the lens space of $L(n,1)$ into a sphere $S^3$.
In Sec.~\ref{sec:summary},  we give the summary and some discussions.

\section{Review}\label{sec:Review}




\subsection{Review of the black lens solution}\label{sec:black lens}

We give a brief review on the supersymmetric black lens solution~\cite{Kunduri:2014kja,Tomizawa:2016kjh} in five-dimensional minimal ungauged supergravity, whose bosonic Lagrangian is described by  the Einstein-Maxwell-Chern-Simons theory: 
\begin{eqnarray}
\mathcal L=R \star 1 -2 F \wedge \star F -\frac 8{3\sqrt 3}A \wedge F \wedge F \,, \label{eq:Lagrangian}
\end{eqnarray}
where $F=d A$ is the Maxwell field.  The metric and gauge potential $1$-form are given, respectively, by
\begin{eqnarray}
\label{metric}
ds^2&=&-f^2(dt+\omega)^2+f^{-1}ds_{M}^2,\\
A&=&\frac{\sqrt 3}{2} \left[f(d t+\omega)-\frac KH(d \psi+\chi)-\xi \right]\,,
\end{eqnarray}
where $ds^2_M$ is the Gibbons-Hawking metric, which can be written in terms of the spherical coordinates and the harmonic function $H$ on $\mathbb E^3$ that has $n$ point sources as
\begin{eqnarray}
ds^2_M&=&H^{-1}(d\psi+\chi)^2+H[dr^2+r^2(d\theta^2+\sin^2\theta d\phi^2)],\\
H&=&\sum_{i=1}^n\frac{h_i}{r_i}:=\frac{n}{r_1}-\sum_{i=2}^n\frac{1}{r_i}, \label{Hdef}\\
\chi&=&\sum_{i=1}^nh_i \frac{z-z_i}{r_i}.
\end{eqnarray}
Here, $r_i:=|{\bm r}-{\bm r_i}|=\sqrt{r^2-2rz_i\cos\theta +z_i^2}$, where the constants $z_i$ and $n$ takes positive integer values.
Furthermore, the function $f^{-1}$ and the $1$-form $\omega$ in the metric are given, respectively, by
\begin{eqnarray}
f^{-1}&=&H^{-1}K^2+L,\label{eq:fi}\\
\omega&=&\omega_\psi(d\psi+\chi)+\hat \omega,
\end{eqnarray}
where the function $\omega_\psi$ and the $1$-form $\hat \omega$ are written as
\begin{eqnarray}
\omega_\psi&=&H^{-2}K^3+\frac{3}{2} H^{-1}KL+M, \\ 
\hat \omega&=&\sum_{i,j=1(i\not=j)}^n\left(h_im_j+\frac{3}{2}k_i l_j \right)\frac{r^2-(z_i+z_j)r\cos\theta+z_iz_j}{z_{ji}r_ir_j}\nonumber\\
&&-\sum_{i=1}^n\left(m_0h_i+\frac{3}{2}l_0k_i\right)\frac{z-z_i}{r_i}+c,
\end{eqnarray}
and  $ K, L, M$ are the harmonic functions with $n$ point sources on $\mathbb E^3$, which are given by
\begin{eqnarray}
K&=&\sum_{i=1}^n\frac{k_i}{r_i},\label{Kdef}\\
L&=&l_0+\sum_{i=1}^n\frac{l_i}{r_i},\label{Ldef}\\
M&=&m_0+\sum_{i=1}^n\frac{m_i}{r_i},\label{Mdef}
\end{eqnarray}
and $c$ is  a constant  and $z_{ji}:=z_j-z_i$. The $1$-form $\xi$ in the gauge potential $A$ is written as
\begin{eqnarray}
\xi&=-&\sum_{i=1}^nk_i \frac{z-z_i}{r_i}.
\end{eqnarray}
The point  source ${\bm r}={\bm r}_1$ corresponds to a degenerate Killing horizon which has the lens space topology $L(n,1)$ and each point source ${\bm r}={\bm r}_i\ (i=2,\ldots,n)$ denotes a regular timelike surface as an origin in the Minkowski spacetime under the appropriate conditions on the parameters. 
\medskip

The solution contains $4n+1$ parameters ($c, l_0, m_0, k_i, l_i, m_{i\ge 2}, z_{i\ge 2}$) but must satisfy the equations
\begin{eqnarray}
l_0&=&1,\label{eq:l0}\\
c&=&-\sum_{i,j(i\not =j)}\frac{h_im_j+\frac{3}{2}k_il_j}{z_{ji}},\label{eq:c}\\
m_0&=&-\frac{3}{2}\sum_{i}k_il_0,\label{eq:m02}
\end{eqnarray}
\begin{eqnarray}
l_i&=&k_i^2\ (i=2,\ldots,n),\label{eq:condition1}\\
m_i&=&\frac{1}{2}k_i^3\ (i=2,\ldots,n),\label{eq:condition2}\\
c_2&:=&m_0-\frac{3}{2}k_i\nonumber \\
&&+\sum_{j(\not =i)}\frac{1}{|z_{ji}|}[3k_i^2k_j+2k_i^3h_j-\frac{3}{2}(k_il_j+l_ik_j+k_il_ih_j)+m_j]=0\ (i=2,\ldots,n)\label{eq:c2}
\end{eqnarray}
and the inequalities
\begin{eqnarray}
R_1^2&:=&k_1^2+nl_1<0,\label{eq:R1ineq}\\
R_2^2&:=&\frac{l_1^2(3k_1^2+4nl_1)}{R_1^4}<0,\label{eq:R2ineq}\\
c_1&:=&1+\sum_{j(\not=i)}\frac{1}{|z_{ji}|}(l_j-2k_ik_j-k_i^2h_j)<0\ (i=2,\ldots,n),\label{eq:c1ineq}
\end{eqnarray}
where we assume $z_i>z_j$ for $i>j$ and  $z_{ij}:=z_i-z_j$.
Equations.~(\ref{eq:l0})---(\ref{eq:m02})  are required from asymptotic flatness and 
Eqs.~(\ref{eq:condition1})---(\ref{eq:c2}) remove curvature singularities and causal violation around each ${\bm r}_i\ (i=2,\ldots,n)$. 
In particular, the conditions (\ref{eq:c2}) are referred to as ``bubble equations'' in Refs.~\cite{Bena:2005va,Bena:2007kg}. 
The inequalities (\ref{eq:R1ineq}) and (\ref{eq:R2ineq}) exclude closed timelike curves (CTCs) around the horizon ${\bm r}_1$, and  the inequality (\ref{eq:c1ineq})  ensures that the spacetime metric is Lorenzian around each ${\bm r}_i\ (i=2,\ldots,n)$. 
Thus, the physical requirements of regularity and causality reduce the number of parameters to $n+1$. 
The Arnowitt-Deser-Misner (ADM) mass and two ADM angular momenta can be computed, respectively, as
\begin{eqnarray}
M&=&\frac{\sqrt{3}}{2}Q=3\pi\left[\left(\sum_{i}k_i\right)^2+\sum_{i}l_i\right],\\
J_\psi&=&4\pi \left[ \left(\sum_{i}k_i\right)^3
+\frac{1}{2}\sum_{i=2}^nk_i^3+\frac{3}{2}\left(\sum_{i}k_i\right)\left(l_1+\sum_{i=2}^nk_i^2\right)\right],\label{eq:jpsi}\\
J_\phi&=&6\pi\left[  \left(\sum_{i}k_i\right) \left(
\sum_{j=2}^nz_j\right)+\left(\sum_{i =2}^nk_iz_i\right)  \right], \label{eq:jphi}
\end{eqnarray}
where $Q$ is the electric charge, which saturates the Bogomol'nyi bound. 
Let $(x,y,z)$ be Euclidean coordinates on ${\mathbb E}^3$ in the Gibbons-Hawking space.
The $z$ axis of ${\mathbb E}^3$ is split into  the $(n+1)$ intervals: $I_-=\{(x,y,z)|x=y=0,  z<z_1\}$, $I_i=\{(x,y,z)|x=y=0,z_i<z<z_{i+1}\}\ (i=1,...,n-1)$, and $I_+=\{(x,y,z)|x=y=0,z>z_n\}$. 
The magnetic fluxes through $I_i\ (i=1,...,n-1)$ are defined as 
\begin{eqnarray}
q[I_i]:=\frac{1}{4\pi}\int_{I_i}F\,,
\end{eqnarray}
which gives
\begin{eqnarray}
q[I_1]=\frac{\sqrt{3}}{2}\left[
\frac{k_1l_1}{2(k_1^2+nl_1)}
-k_2
\right]\,, \qquad 
q[I_i]=\frac{\sqrt{3}}{2}(k_i-k_{i+1})~~ (i=2,... n-1).
\label{mag_fluxes}
\end{eqnarray}
In particular, for $n=1$, this solution recovers the BMPV black hole solution.




\subsection{Review of the Klemm-Sabra solution}\label{sec:Klemm}
Next, we briefly review the Klemm-Sabra solution from Ref.~\cite{Klemm:2000gh}, which can be regarded as the BMPV black hole \cite{Breckenridge:1996is} in de Sitter or anti-de Sitter spacetime in the EMCS theory with a cosmological constant, whose Lagrangian is given by
\begin{eqnarray}
\mathcal L=(R+2\Lambda) \star 1 -2 F \wedge \star F -\frac 8{3\sqrt 3}A \wedge F \wedge F \,. \label{eq:Lagrangian2}
\end{eqnarray}
In particular, for $\Lambda>0$, the solution can be expressed in the cosmological coordinates as follows:
\begin{eqnarray}
 ds^2 &=&- \left( \lambda \tau + \frac{m}{\rho^2} \right)^{-2} 
  \left[ d\tau + \frac{j}{2 \rho^2} \left( d\psi + \cos \theta d\phi \right) \right] ^2 
 \notag \\
        &&+ \left( \lambda \tau + \frac{m}{\rho^2} \right)
  \left[ d\rho^2 + \frac{\rho^2}{4} \left\{ d\theta^2+\sin^2\theta d\phi^2 
  +\left( d\psi + \cos \theta d\phi \right)^2 \right\} \right], \label{KSmet} \\
  A&=&\frac{\sqrt{3}}{2}\left(\lambda\tau+\frac{m}{\rho^2} \right)^{-1}\left[ d\tau+ \frac{j}{2\rho^2}(d\psi+\cos\theta d\phi)\right].\label{KSA}
\end{eqnarray}
The constants $m$ and $j$ are the mass parameter and angular momentum parameter, respectively and the constant $\lambda$ is related to the positive cosmological constant  by $\lambda=\pm2\sqrt{\Lambda/3}$.
This metric seems to have time dependence due to the existence of the time coordinate $\tau $ in the metric, but it can be shown that it has a stationary region (see Ref.~\cite{Matsuno:2007ts}). 

\medskip
To see the locations of the apparent horizons for the Klemm-Sabra black hole spacetime, let us define $x:=\lambda\tau \rho^2$.
 In terms of this $x$, the expansions of the outgoing and ingoing null geodesic congruences for $\psi,\phi,\theta={\rm constant}$ can be computed as
\begin{eqnarray}
\theta_\pm=\lambda\pm\frac{2x}{\sqrt{(x+m)^3-j^2}}.
\end{eqnarray}
Hence, it turns out that three horizons exist at $x=x_\pm,x_c$ $(x_-<x_+<x_c)$, which are the three roots of the cubic equation
\begin{eqnarray}
\lambda^2[(x+m)^3-j^2]-4x^2=0 \label{eq:cubic}
\end{eqnarray}
when the two parameters satisfy the inequalities
\begin{eqnarray}
 0\le m \lambda^2 \le \frac{2}{3}, \quad 
 j_-^2 (m) \le j^2 \le j_+^2 (m), \label{hanni}
\end{eqnarray}
where  
\begin{eqnarray} 
 j_\pm^2 (m) = \frac{4}{27 \lambda ^6} \left[9 m\lambda ^2 (8-3 m\lambda ^2) -32 \pm 8
   \sqrt{2} (2-3 m\lambda ^2 )^{3/2}\right]. 
\end{eqnarray}
In the case of $j=j_+$, the black hole horizon $x_+$ coincides with  
the inner horizon $x_-$, and in the case of $j=j_-$, the black hole horizon $x_+$ coincides with  the cosmological horizon $x_c$.  
The naked singularity appears if $m$ and $j$ are outside of the ranges (\ref{hanni}).
The curvature singularity exists at $x=-m$.   
Moreover, we can show the absence of CTCs on/outside the black hole horizon since 
the two-dimensional $(\psi,\phi)$ part of the metric can be shown to be positive within the ranges (\ref{hanni}). 
Finally, for use in the next section, we should keep in mind that this BMPV solution can be formally obtained by replacing $  ``\lambda\tau"$ of $\lambda\tau+m/\rho^2$ in Eqs.~(\ref{KSmet}) and (\ref{KSA}) with the constant $ ``1."$




\section{Cosmological black lens solution}
\label{sec:solution}

To obtain a cosmological black lens solution, we consider the special case of $k_i=\alpha h_i$ \ (where $\alpha$ is a constant) for $i=1,\ldots,n$ in the supersymmetric black lens solution, namely, 
\begin{eqnarray} 
K=\alpha H. \label{eq:assume}
\end{eqnarray}
Then, the conditions~(\ref{eq:c})-(\ref{eq:c2}) are written in simpler forms, respectively,  as
\begin{eqnarray}
c&=&-\sum_{i,j(i\not =j)}\frac{h_i(m_j+\frac{3}{2}\alpha l_j)}{z_{ji}},\label{eq:ca}\\
m_0&=&-\frac{3}{2}\alpha, \label{eq:m0a}\\
l_i&=&\alpha^2\ (i=2,\ldots,n) ,\label{eq:condition1a}\\
m_i&=&-\frac{1}{2}\alpha^3\ (i=2,\ldots,n),\label{eq:condition2a}\\
l_1&=&-\frac{2}{3}n\alpha^2. \label{eq:c2a}
\end{eqnarray}
Also, the inequities (\ref{eq:R1ineq}),  (\ref{eq:R2ineq}), and  (\ref{eq:c1ineq}) can be written, respectively,  as
\begin{eqnarray}
R_1^2&:=&\alpha^2n^2+nl_1>0,\label{eq:R1ineqa}\\
R_2^2&:=&\frac{l_1^2(3\alpha^2n^2+4nl_1)}{R_1^4}>0,\label{eq:R2ineqa}\\
(c_1&:=&)\ 1+\frac{l_1+n\alpha^2}{z_{i1}}<0.\label{eq:c1ineqa}
\end{eqnarray}
We should note that when the condition (\ref{eq:c2a}) holds, both of the inequalities (\ref{eq:R1ineqa}) and   (\ref{eq:R2ineqa}) are automatically satisfied, whereas the inequality (\ref{eq:c1ineqa}) cannot be satisfied since the left-hand side is positive. 
Therefore,  the metric never becomes Lorenzian near each point of ${\bm r}={\bm r}_i\ (i=2,\ldots,n)$ under the parameter setting (\ref{eq:assume}). 
This fact  implies that a physical black lens with two zero angular momenta cannot be realized since it can be shown from Eq.~(\ref{eq:c2a}) that two angular momenta [Eqs.(\ref{eq:jpsi}) and (\ref{eq:jphi})] vanish when Eq.(\ref{eq:assume}) is imposed. 
As will be seen later, however, the existence of a positive cosmological constant changes this situation since a negative term relating to the cosmological constant term appears on the left-hand side. 
We will show that at least, at late time and early time, the parameters are such that all of these conditions can be satisfied.

\medskip
Let us recall that the Klemm-Sabra solution~(\ref{KSmet}) expressed in terms of the cosmological coordinates can be obtained by replacing the constant $1$ in the harmonic function $f=1+m/\rho^2$ with $\lambda \tau$ for the BMPV solution,  where $\tau$ is a cosmological time coordinate and $\lambda=\pm2\sqrt{\Lambda/3}$. 
In the same way, let us formally replace the constant $l_0$ in Eq.(\ref{eq:fi}) with $\lambda \tau$. 
Then, we can see that the obtained metric and gauge potential 1-form are the solutions in the five-dimensional EMCS theory with a positive cosmological constant, whose Lagrangian is given by Eq.~(\ref{eq:Lagrangian2}). 

\medskip

The metric and gauge potential obtained here for the cosmological black lens are presented, respectively,  as
\begin{eqnarray}
\label{metric}
ds^2&=&-f^2(d\tau+\omega)^2+f^{-1}ds_{M}^2,\\
A&=&\frac{\sqrt 3}{2} \left[f(d \tau+\omega)-\alpha (d \psi+\chi)-\xi \right]\,,
\end{eqnarray}
where $ds^2_M$ is  the metric of the Gibbons-Hawking space
\begin{eqnarray}
ds^2_M&=&H^{-1}(d\psi+\chi)^2+H[dr^2+r^2(d\theta^2+\sin^2\theta d\phi^2)],\\
H&=&\sum_{i=1}^n\frac{h_i}{r_i}:=\frac{n}{r_1}-\sum_{i=2}^n\frac{1}{r_i}, \label{Hdef}\\
\chi&=&\sum_{i=1}^nh_i \frac{z-z_i}{r_i}.
\end{eqnarray}
The function $f^{-1} $ and the  $1$-form $\omega$ are written as
\begin{eqnarray}
f^{-1}
         &=&\lambda \tau +\sum_i\frac{l_i+\alpha^2 h_i}{r_i},\\
\omega&=&\omega_\psi(d\psi+\chi)+\hat \omega.
\end{eqnarray}
 The function $\omega_\psi$ and $1$-forms $(\hat \omega,\xi)$ are given, respectively,  by
\begin{eqnarray}
\omega_\psi                   &=&m_0+\frac{3}{2}\alpha l_0+\sum_{i}\frac{\alpha^3h_i+\frac{3}{2}\alpha l_i+m_i}{r_i},
\end{eqnarray}
\begin{eqnarray}
\hat \omega&=&\Biggl[\sum_{i,j=1(i\not=j)}^nh_i\left(m_j+\frac{3}{2}\alpha l_j \right)\frac{r^2-(z_i+z_j)r\cos\theta+z_iz_j}{z_{ji}r_ir_j} \notag \\
&& -\sum_{i=1}^nh_i\left(m_0+\frac{3}{2}l_0\alpha\right)\frac{r\cos\theta-z_i}{r_i}+c\Biggl]d\phi \,, 
\end{eqnarray}
\begin{eqnarray}
\xi&=-&\alpha\sum_{i=1}^nh_i \frac{z-z_i}{r_i}.
\end{eqnarray}
This solution has $(3n+4)$ constants $(l_0,l_{i\ge 1},m_0,m_{i\ge1},z_{i\ge1},\alpha,c)$,  
but as will be explained later, in order that we can regard this as a physical solution, appropriate conditions must be imposed on these parameters.




\section{analysis}
\label{sec:analysis}
In this section, we analyze the cosmological black lens solution obtained in the previous section, focusing on the the contracting phase $\lambda<0$.
Using the cosmological coordinates $\tau$,  we can see that the spatial cross section of the apparent horizon changes from the lens space $L(n,1)=S^3/{\mathbb Z}_n$ into a sphere $S^3$.  It is a hard task to see the process of this topology change analytically, but it is easy to see the asymptotic behaviors of  the solution at early time $\tau=-\infty$ and late time $\tau=0$, since the spacetime near ${\bm r}={\bm r}_1$ and $r=\infty$ becomes (locally) the Klemm-Sabra spacetime, namely, it asymptotically becomes stationary.




\subsection{Early time}
First of all, we show that the spacetime around  $r:=|\bm r-\bm r_1|=0$ can be locally approximated by the Klemm-Sabra black hole spacetime, namely, the spacetime asymptotically becomes stationary, at least, around this region. In fact, it turns out that 
for $r=|{\bm r}-{\bm r}_1| \to 0$, 
the metric behaves as
\begin{eqnarray}
ds^2&\simeq&- \left(\lambda\tau +\frac{l_1+n\alpha}{r}\right)^{-2}\left[d\tau+\left(\frac{n\alpha^3+\frac{3}{2}\alpha l_1+m_1}{r}\right)(d\psi+n\cos\theta d\phi)\right]^2 \notag\\
&+&\left(\lambda\tau +\frac{l_1+n\alpha}{r}\right)\left[\frac{r}{n}(d\psi+n\cos\theta d\phi)^2+\frac{n}{r}\left(dr^2+r^2(d\theta^2+\sin^2\theta d\phi^2)\right)\right],
\end{eqnarray}
where we have shifted $\tau$ appropriately. 
In terms of the five-dimensional radial coordinate $\rho:=\sqrt{4nr}$, the asymptotic form of the metric can be rewritten as
\begin{eqnarray}
ds^2&\simeq &- \left(\lambda\tau +4n\frac{l_1+n\alpha}{\rho^2}\right)^{-2}\left[d\tau+\left(4n^2\frac{n\alpha^3+\frac{3}{2}\alpha l_1+m_1}{\rho^2}\right)\left(\frac{d\psi}{n}+\cos\theta d\phi \right)\right]^2 \notag\\
&+&\left(\lambda\tau +4n\frac{l_1+n\alpha}{\rho^2}\right)\left[d\rho^2+\frac{\rho^2}{4}\left\{\left(\frac{d\psi}{n}+\cos\theta d\phi\right)^2+d\theta^2+\sin^2\theta d\phi^2\right\}\right].\label{eq:latetime}
\end{eqnarray}
We observe from Eq.(\ref{KSmet}) that this metric is locally isometric to that of the Klemm-Sabra solution obtained by identifying $(m,j)$ in Eq.~(\ref{KSmet}) with $(4n(l_1+n\alpha),8n^2(n\alpha^3+\frac{3}{2}n\alpha l_1+m_1))$.  
Therefore, the sufficiently small closed surface that is centered at $r=0\ ({\bm r}={\bm r}_1)$ turns out to be outer trapped at early time $\tau=-\infty$ since $\theta_+=0$ holds at $\rho^2=x_+(4n(l_1+n\alpha),8n^2(n\alpha^3+\frac{3}{2}n\alpha l_1+m_1))/(\lambda\tau)$ for a sufficiently small $\tau$. 
From Eq.~(\ref{eq:lens}), we must note that the spatial cross section of the apparent horizon is topologically not a sphere $S^3$, but rather the lens space of $L(n,1)=S^3/{\mathbb Z}_n$, since $\psi$ in the asymptotic metric~(\ref{eq:latetime})  is divided by $n$.

\medskip
Second, we show that the spacetime near each $\bm r=\bm r_i\ (i=2,\dots,n)$ behaves as an origin in the Minkowski spacetime written in polar coordinates.  
For $r:=|{\bm r}-{\bm r}_i| \to 0$ $(i=2,\ldots,n)$ and $\tau\to-\infty $ and keeping $\lambda \tau r=:\beta$\ (where $\beta$ is a positive constant), 
the functions $(f^{-1},\omega_\psi)$ are approximated as
\begin{eqnarray}
f^{-1}&\simeq& \frac{l_i+\alpha^2h_i+\beta}{r}+c_1,\\
\omega_\psi &\simeq& \frac{\alpha^3h_i+\frac{3}{2}\alpha l_i+m_i}{r}+c_2,
\end{eqnarray}
where the constants $c_1$ and $c_2$ are defined by
\begin{eqnarray}
c_1&:=& \sum_{j(\not=i)}\frac{l_j+\alpha^2 h_j}{|z_{ji}|},\\
c_2&:=& m_0+\frac{3}{2}\alpha l_0+\sum_{j(\not=i)}\frac{\alpha^3h_j+\frac{3}{2}\alpha l_j+m_j}{|z_{ji}|}.
\end{eqnarray}
The metric behaves as
\begin{eqnarray}
ds^2&\simeq& -\left(\frac{l_i+\alpha^2h_i+\beta}{r}+c_1\right)^{-2}
\biggl[d\tau+\left(\frac{\alpha^3h_i+\frac{3}{2}\alpha l_i+m_i}{r}+c_2\right)\nonumber\\
&\times&  \left\{d\psi+(-\cos\theta +\chi_{(0)})d\phi\right\}+(\hat\omega_{(1)}\cos\theta+\hat \omega_{(0)})d\phi \biggr]^2\nonumber\\
&-&\left(\frac{l_i+\alpha^2h_i+\beta}{r}+c_1\right)r\left[\left\{d\psi+(-\cos\theta +\chi_{(0)})d\phi\right\}^2+\frac{dr^2}{r^2}+d\theta^2+\sin^2\theta d\phi^2\right],
\end{eqnarray}
where $\chi_{(0)}$, $\hat\omega_{(0)}$, and $\hat\omega_{(1)}$ are given, respectively,  by
\begin{eqnarray}
\chi_{(0)} & := &- \sum_{j(\not=i)}\frac{h_jz_{ji}}{|z_{ji}|} \,,\\
\hat \omega_{(0)}
& := &   \sum_{k,j(\not=i,k\not=j)}\left(h_km_j+\frac{3}{2}\alpha h_kl_j\right)\frac{z_{ji}z_{ki}}{|z_{ji}z_{ki}|z_{jk}}+\sum_{j(\not= i)}\left(m_0h_j+\frac{3}{2}\alpha h_jl_0\right)\frac{z_{ji}}{|z_{ji}|}+c\,, \\
\hat \omega_{(1)}& := & -\sum_{j(\not=i)}\left(h_im_j-h_jm_i+\frac{3}{2}\alpha(h_il_j-h_jl_i)\right)\frac{1}{|z_{ji}|}-\left(m_0h_i+\frac{3}{2}\alpha h_il_0\right)\,.
\end{eqnarray}

\medskip
To remove the divergence of the metric at $r=|{\bm r}-{\bm r}_i|=0\ (i=2,\ldots,n)$, we impose the following conditions on the set of parameters $(l_i,m_i)\ (i=2,\ldots,n)$:
\begin{eqnarray}
l_i&=&\alpha^2-\beta, \label{eq:li}\\
m_i&=&-\frac{1}{2}\alpha^3+\frac{3}{2}\alpha \beta,\label{eq:mi}
\end{eqnarray}
which imply
\begin{eqnarray}
c_2=\hat\omega_{(1)}.
\end{eqnarray}
Therefore, the metric is
\begin{eqnarray}
ds^2&\simeq& -c_1^{-2}\biggl[d\tau +c_2\left\{d\psi+\chi_{(0)}d\phi\right\}+\hat \omega_{(0)}d\phi \biggr]^2\\
&-&c_1r\left[\frac{dr^2}{r^2}+\left\{d\psi+(-\cos\theta +\chi_{(0)})d\phi\right\}^2+d\theta^2+\sin^2\theta d\phi^2\right].
\end{eqnarray}
Since the existence of $c_2$ and $\hat\omega_{(0)}$ yield CTCs around ${\bm r}={\bm r}_i\ (i=2,\ldots,n)$, 
furthermore, we must impose the following additional condition
\begin{eqnarray}
(c_2=)\  m_0+\frac{3}{2}\alpha l_0+\frac{n\alpha^3+\frac{3}{2}l_1\alpha+m_1}{z_{i1}}=0. \label{eq:c20}
\end{eqnarray}
It can be shown from $c_2=0$ that $\hat \omega_{(0)}=0$ automatically holds.
In addition, in order that the metric has Lorenzian signature, we need to require  
\begin{eqnarray}
(c_1=)\ \frac{l_1+n\alpha^2}{z_{i1}}-\beta\sum_{2\le j(\not=i)}\frac{1}{|z_{ji}|}<0. \label{eq:c1ineq2}
\end{eqnarray}
Under this parameter setting, the asymptotic form of the metric can be written as
\begin{eqnarray}
ds^2
&=&-(dt')^2+d\rho^2+\frac{\rho^2}{4}\left[(d\psi'-\cos\theta d\phi)^2+d\theta^2+\sin^2\theta d\phi^2\right],
\end{eqnarray}
where we have introduced the new coordinates $(t',\psi',\rho)$
\begin{eqnarray}
t'=c_1^{-1}\tau,\quad \psi'=\psi+\chi_{(0)}\phi,\quad \rho=2\sqrt{-c_1r}.
\end{eqnarray}
As we showed in Sec.II,  for the stationary black lens solution with $k_i=\alpha h_i\ (i=1,\ldots, n)$, the inequality (\ref{eq:c1ineq2}) with $\beta=0 $ cannot be satisfied.
In contrast to this, for the cosmological black lens solution, the parameter region satisfying the inequality (\ref{eq:c1ineq2}) indeed exists since the first term on the left-hand side of Eq.(\ref{eq:c1ineq2}) is still positive, while the second term---which obviously appears by the existence of a positive cosmological constant---is negative. 
Hence, under these conditions, the region around ${\bm r} ={\bm r}_i\ (i=2,\ldots,n)$ at early time is  isometric to the Minkowski spacetime.




\subsection{Late time}
For $\rho:=2\sqrt{r}\to \infty$, under the conditions~(\ref{eq:li}) and (\ref{eq:mi}), the two functions $(f^{-1},\omega_\psi)$ and the $1$-form $\hat \omega$ behave, respectively, as 
\begin{eqnarray}
f^{-1}
&\simeq&\lambda\tau +\frac{l_1+n\alpha^2-(n-1)\beta}{r}+{\rm const.},\label{eq:fi}\\
\omega_\psi&\simeq& m_0+\frac{3}{2}\alpha l_0+\frac{\alpha^3\sum_ih_i+\frac{3}{2}\alpha \sum_il_i+\sum_im_i}{r}\\
                     &=&\frac{\alpha^3+\frac{3}{2}\alpha l_1+m_1}{r},
\end{eqnarray}
and
\begin{eqnarray}
\hat \omega_\phi&\simeq &\sum_{i,j=1(i\not=j)}^n\left(h_im_j+\frac{3}{2}k_i l_j \right)\frac{1}{z_{ji}}\left(1-\frac{z_{ji}^2\sin^2\theta}{2r^2}+{\cal O}(r^{-3})\right) \notag \\
&& -\sum_{i=1}^n\left(m_0h_i+\frac{3}{2}l_0k_i\right)\left(\cos\theta -\frac{z_i\sin^2\theta}{r}+{\cal O}(r^{-2})\right)+c\\
&=& \sum_{i,j=1(i\not=j)}^n\left(h_im_j+\frac{3}{2}k_i l_j \right)\frac{1}{z_{ji}}+c  -\sum_{i=1}^n\left(m_0h_i+\frac{3}{2}l_0k_i\right)\cos\theta  \notag \\
&& -\sum_{i=1}^n\left(m_0h_i+\frac{3}{2}l_0k_i\right)\frac{-z_i\sin^2\theta}{r}+{\cal O}(r^{-2})\\
&=& \sum_{i=1}^n\left(m_0+\frac{3}{2}l_0\alpha\right)h_i\left(\cos\theta-\frac{-z_i\sin^2\theta}{r}\right)+{\cal O}(r^{-2})\\
&=&{\cal O}(r^{-2}),
\end{eqnarray}
where to ensure an asymptotic de Sitter space (or, the absence of CTCs) at $r\to \infty$,  we have imposed
\begin{eqnarray}
c&=&-\sum_{i,j}\left(h_im_j+\frac{3}{2}k_il_j\right)\frac{1}{z_{ij}},\\
m_0&+&\frac{3}{2}\alpha l_0=0. \label{eq:m0}
\end{eqnarray}
\medskip

Then, the metric can be approximately written as
\begin{eqnarray}
ds^2&\simeq&- \left(\lambda\tau +\frac{l_1+n\alpha^2-(n-1)\beta}{r}\right)^{-2}\left[d\tau+\frac{\alpha^3+\frac{3}{2}\alpha l_1+m_1
}{r}(d\psi+\cos\theta d\phi)\right]^2 \notag\\
&+& \left(\lambda\tau +\frac{l_1+n\alpha^2-(n-1)\beta}{r}\right)\left[r(d\psi+\cos\theta d\phi)^2+\frac{1}{r}\left(dr^2+r^2d\Omega_{S^2}^2\right)\right]\\
&=&-  \left(\lambda\tau +\frac{4[l_1+n\alpha^2-(n-1)\beta]}{\rho^2}\right)^{-2}\left[d\tau+\frac{4\{\alpha^3+\frac{3}{2}\alpha l_1+m_1
\}}{\rho^2}(d\psi+\cos\theta d\phi)\right]^2\notag\\
&+&\left(\lambda\tau +\frac{4[l_1+n\alpha^2-(n-1)\beta]}{\rho^2}\right)\left[d\rho^2+\frac{\rho^2}{4}\left\{(d\psi+\cos\theta d\phi)^2+d\theta^2+\sin^2\theta d\phi^2\right\}\right],
\end{eqnarray}
where we have shifted $\tau$ so that the constant term in~Eq.(\ref{eq:fi}) vanishes. 
Compared with Eq.~(\ref{KSmet}), we immediately find that this obtained asymptotic metric coincides with the metric of the Klemm-Sabra solution with the mass parameter $m=4[l_1+n\alpha^2-(n-1)\beta]$ and angular momentum parameter $j=8[\alpha^3+\frac{3}{2}\alpha l_1+m_1]$. 
Therefore, a sufficiently large closed surface $r={\rm const}$ turns out to be outer trapped at late time $\tau\simeq -0$ since $\theta_+=0$ holds at $\rho^2=x_+(4(l_1+n\alpha^2-(n-1)\beta),8(\alpha^3+\frac{3}{2}\alpha l_1+m_1))/(\lambda\tau)$ if $\tau$ is sufficiently close to $0$ and negative. 
Here, we must note that the spatial cross section of the apparent horizon is topologically a sphere $S^3$, whereas it is topologically the lens space $L(n,1)=S^3/{\mathbb Z}_n$ at early time. 

\medskip
From the behaviors at early time and at late time,, we can make the physical interpretation that the solution obtained here describes the topology change of a black hole from the lens space $L(n,1)$ into a sphere $S^3$.
We note from the conditions~(\ref{eq:m0}) and (\ref{eq:c20}) that at early time, the angular momentum along $\partial/\partial\psi$, $j$ vanishes, whereas at late time it does not vanish.  
Hence, apparently, the angular momentum seems to not be preserved but  this is not so very surprising since at early time, the Maxwell field exists outside the black hole horizon, which can be considered to contribute to the mass and angular momenta.




\subsection{$\tau=$const.}
It seems to be difficult to analytically see the behavior of the solution for $-\infty<\tau<0$ since the spacetime is not stationary. 
However, we can see the behaviors around the points ${\bm r}={\bm r}_i\ (i=2,\ldots,n)$ at which the metric diverges for finite $\tau\ (-\infty<\tau<0)$.
For $r:=|{\bm r}-{\bm r}_i| \to 0$ and  finite $\tau\ (-\infty<\tau<0)$, the metric functions $(f^{-1},\omega_\psi)$ behave as
\begin{eqnarray}
f^{-1}&\simeq&
\frac{-\beta}{r}+\lambda\tau+c_1,\\
\omega_\psi &\simeq& \frac{\alpha^3h_i+\frac{3}{2}\alpha l_i+m_i}{r}+c_2={\cal O}(r) ,\\
\hat \omega_\phi&\simeq& \hat \omega_{(1)}\cos\theta+\hat\omega_{(0)}= 0,
\end{eqnarray}
where we have used the conditions~(\ref{eq:li}), (\ref{eq:mi}), and (\ref{eq:c20}). 
The metric is approximated by
\begin{eqnarray}
ds^2&\simeq&\frac{-r^2}{\beta^2}dt^2+d\rho^2+\beta \left[(d\psi'-\cos\theta d\phi)^2+d\theta^2+\sin^2\theta d\phi^2 \right],
\end{eqnarray}
 where we have introduced the new coordinates $(\psi',\rho)$
\begin{eqnarray}
\psi'=\psi+\chi_{(0)}\phi,\quad \rho=\sqrt{\beta}\log r.
\end{eqnarray}
Thus, the divergence of the metric at $r=0$ has been removed. 
It turns out from the above asymptotic form that for $-\infty<\tau<0$, each point ${\bm r}={\bm r}_i$ approximately behaves like a Killing horizon with the spatial cross section of a sphere of radius $2\sqrt{\beta}$.




\section{Summary}
\label{sec:summary}
In this paper, we have obtained the cosmological black lens solution in the five-dimensional EM theory with a Chern-Simons term and a positive cosmological constant. 
We have also discussed some properties of the rotating charged black lens solution in terms of the cosmological coordinates. 
 It has been shown that this solution can be regarded as a dynamical black hole spacetime such that at early time, the horizon cross section is isometric to the lens space $L(n,1)=S^3/{\mathbb Z}_n$, while at late time, it is isometric to a sphere $S^3$. 
This solution has been obtained from the special limit of the extreme black lens in Refs.~\cite{Kunduri:2014kja,Tomizawa:2016kjh} in the same way that Klemm and Sabra's cosmological charged black hole was obtained from the BMPV black hole solution.
In this restricted limit, the supersymmetric black lens spacetime behaves pathologically near the point sources $\bm r=\bm r_i\ (i=2,\ldots,n)$ outside the horizon, whereas the cosmological solution does not show this behavior (at least, at early time and late time) due to the presence of a cosmological constant.   

\medskip 
For $n=1$, our solution exactly coincides with the Klemm-Sabra solution that can be physically regraded as a stationary, rotating black hole in de Sitter space. 
It hence follows that the angular momentum along $\partial/\partial \psi$ is conserved. 
For $n\ge 2$, in turn, it apparently seems to describe the dynamical spacetime that violates the conservation of angular momentum. 
In fact,  it can be seen from Eqs.~(\ref{eq:c20}) and (\ref{eq:m0}) that at early time, ${\bm r}={\bm r}_1$ locally behaves like the five-dimensional Reissner-Nordstr\"om de Sitter black hole ({\it i.e.,} the Klemm-Sabra black hole with zero angular momentum), whereas at late time, $r=\infty$ behaves as the Klemm-Sabra black hole with nonzero angular momentum.  
We may consider that this is because at early time, the Maxwell field outside the black hole horizon carries the same amount of the angular momentum.

\medskip

 The  cosmological multi-black-hole solutions were first obtained by Kastor and Trashen~\cite{Kastor:1992nn} for the four-dimensional EM theory with a positive cosmological constant. 
 Furthermore, these solutions were immediately generalized to the five-dimensional EM theory with a cosmological constant and a Chern-Simons term  in Ref.~\cite{London:1995ib} for the nonrotating case and in Ref.~\cite{Klemm:2000gh} for the rotating case, respectively.
For the present solution, we can consider the multicentered black hole solution if one does not impose the conditions (\ref{eq:li}), (\ref{eq:mi}), and (\ref{eq:c20}). 
It is easily expected that such a solution describes the coalescence of black lenses since multiple black holes with various lens space topologies exist at early time, while only a single black hole with spherical topology exists at late time.

\medskip
 From the viewpoint of the AdS/CFT correspondence, one of the most interesting generalizations is to look for an anti-de Sitter black lens.  
Concerning the spherical topology,  the charged rotating black hole solutions were obtained in the five-dimensional EMCS theory with a negative cosmological constant in Refs.~\cite{Klemm:2000gh,Gutowski:2004ez,Gutowski:2004yv,Chong:2004na,Chong:2005da,Chong:2005hr,Chong:2006zx}. 
As for the ring topology $S^1\times S^2$, it was shown that there is no supersymmetric black ring solution in five-dimensional minimal gauged supergravity, although it is not yet known if one exists in nonminimal supergravity.    
These results seem to not prohibit the existence of a black lens solution even in five-dimensional minimal gauged supergravity.   
Therefore, to construct such solutions is also an interesting and challenging problem.
 This issue deserves further study.




\acknowledgments

This work was supported by the Grant-in-Aid for Scientific Research (C) (Grant Number ~17K05452) from the Japan Society for the Promotion of Science (S.T.).








\end{document}